\documentclass[preprint,amsmath,amssymb,nofootinbib]{revtex4}

\usepackage{color}
\usepackage{graphicx}% Include figure files

\usepackage{dcolumn}% Align table columns on decimal point
\usepackage{bm}% bold math
\usepackage{lscape}
\usepackage{ulem}

\usepackage{txfonts}

\begin{document}
\title{Relation between transition density and proton inelastic scattering\\
by $^{12}$C target at $E_p =$ 65 and 200 MeV}

\author{T.~Furumoto}
\email{furumoto-takenori-py@ynu.ac.jp}
\affiliation{Graduate School of Education, Yokohama National University, Yokohama 240-8501, Japan}

\author{M.~Takashina}%
\affiliation{Osaka Heavy Ion Therapy Center, Osaka 540-0008, Japan}

\date{\today}

\begin{abstract}
We calculate proton elastic and inelastic scatterings with a microscopic coupled channel (MCC) calculation.
The localized diagonal and coupling potentials including the spin-orbit part are obtained by folding a complex $G$-matrix effective nucleon-nucleon interaction with a transition density.
This is the first time that the present folding prescription for the spin-orbit part is applied to the proton inelastic scattering, while for the monopole transition only.
We apply the MCC calculation to the proton elastic and inelastic (0$^+_2$) scatterings by $^{12}$C target at $E_p$ = 65 and 200 MeV.
The role of diagonal and coupling potentials for the central and spin-orbit parts is checked.
In addition, the relation between the transition density and the proton inelastic scattering is investigated with the modified wave function and the modified transition density.
Namely, we perform the investigation with the artificial drastic change rather than fine structural change.
The inelastic cross section is sensitive to the strength and shape of the transition density, but the inelastic analyzing power is sensitive only to the shape of that.
Finally, we make clear the property of the inelastic analyzing power derived from the transition density without an ambiguity.
\end{abstract}

\maketitle

%%%%%%%%%%%%%%%%%%%%%%%%%%%%%%%%%%%%%%%%%%%%%%%%%%%%%%%%%%%%%%%%%%%
% introduction
%%%%%%%%%%%%%%%%%%%%%%%%%%%%%%%%%%%%%%%%%%%%%%%%%%%%%%%%%%%%%%%%%%%
\section{Introduction}
In the universe, nuclear reactions are occurring every day.
Such nuclear reactions give us various heavy elements.
One of key reactions is well known to be carbon synthesis reaction.
The $^8$Be resonance state is constructed by colliding two $\alpha$ particles.
When another $\alpha$ particle is synthesized to the $^8$Be resonance state, they will be excited state of the $^{12}$C nucleus.
By emitting a $\gamma$-ray, the carbon nucleus settles into a stable ground state.
In the reaction, not only the $^8$Be resonance state but also the ground and excited states of $^{12}$C, especially for the 0$_2^+$ (Hoyle) state, has an important role to give the abundance ratio of elements.
Therefore, a lots of researchers investigate the $^{12}$C nucleus from the microscopic viewpoints of nuclear structure and reaction~\cite{UEG77,KAM81,FUN03,FUN06,CHE07,YAM08,EPE12,SUH15,SAK84,SAK86,TAK06,TAK08,KHO11,CUO13,KAN19,VIT20}.

One of key issues for the investigation of the $^{12}$C nucleus is to identify the size of the Hoyle state with the experimental data.
In order to identify the size of the excited state, the $\alpha$ + $^{12}$C reaction data is investigated and discussed~\cite{OHK04,TAK06,TAK08}.
Especially, there was focused on the relation between the angular distribution of the inelastic cross section and the size of excited (0$^+_2$) state.
However, in conclusion, the $\alpha$ inelastic cross section mainly gives us the information of the transition density between the ground state and excited state rather than the size of the excited state~\cite{TAK06,TAK08}.
The relation between the $\alpha$ inelastic cross section and the transition density including the contribution from the size of the excited state is still discussed~\cite{ITO18,KAN19,VIT20}.

In this paper, we investigate the relation between the transition density and the inelastic cross section, but for $p + ^{12}$C reaction.
It is valuable to understand the property of the transition density obtained from not only the incident $\alpha$ particle but also incident other ones.
In addition, the merit of proton scatterings is to give us spin observables.
The spin observables are also a powerful tool to investigate the property of the transition density.
In Ref.~\cite{IID12}, the relation between the inelastic cross section and the size of the excited state is investigated with the black sphere model for the proton scattering.
However, the spin observable is not investigated in the black sphere model.
Although not the Hoyle state, the relation between spin observables and the excited states has been investigated~\cite{DNR,SWI72,LIL79,GUS82}.
However, the investigations have been performed with the phenomenological way in the past.
We have no ambiguity to link the inelastic scatterings and the transition density in the present folding prescription.

In order to describe the proton scatterings, we apply the complex $G$-matrix interaction to the microscopic coupled-channel (MCC) calculation.
We here mention that the localized diagonal and coupling potentials based on the folding procedure are applied to the MCC calculation.
Historically, the microscopic reaction calculation for the proton scatterings is powerfully developed in the formalism of a non-local approach with the single particle wave function~\cite{DER83,HYN85,KAR95,MEL00}.
Such explicit description called as full-folding approach well reproduces a lots of proton elastic scatterings.
However, it is difficult to construct a flawless formalism, especially for the inelastic scatterings.
Namely, their approach is a powerful tool to describe the inelastic scatterings but it is difficult to perform a step-by-step analysis through the transition density and the localized potential.
On the other hand, the development of the localized approach for the proton scatterings is advanced, and the validity of that is investigated~\cite{RIK842,KHO02,CEG07,MIN10}.
In addition, the purpose in this paper is to reveal the relation between the transition density and the proton scatterings.
The microscopic construction of the transition density is one of key issues to investigate the property of the nuclear structure through the nuclear reaction~\cite{TAK05,TAK08-2,FUR13-2,FUR18,SAT19}.
Therefore, it is valuable to describe the proton elastic and inelastic scatterings with the transition density in the localized microscopic framework.

In this paper, we first introduce the folding prescription for the central and spin-orbit parts of the diagonal and coupling potentials in Sec.~\ref{sec:formalism}.
In Sec.~\ref{sec:results}, we will show the results of the present MCC calculation for the $p + ^{12}$C scatterings at 65 and 200 MeV.
The effect of the central and spin-orbit parts of the diagonal and coupling potentials is clarified on the inelastic cross section and analyzing power.
The important role of the transition density is also revealed in inelastic cross section and analyzing power.
Lastly, we will summarize this paper in Sec.~\ref{sec:summary}.

%\vspace{2mm}
%%%%%%%%%%%%%%%%%%%%%%%%%%%%%%%%%%%%%%%%%%%%%%%%%%%%%%%%%%%%%%%%%%%
% formalism
%%%%%%%%%%%%%%%%%%%%%%%%%%%%%%%%%%%%%%%%%%%%%%%%%%%%%%%%%%%%%%%%%%%
\section{Formalism}
\label{sec:formalism}

To solve the coupled-channel equation, the diagonal and coupling potentials are needed.
In the present microscopic approach, the localized diagonal and coupling potentials are obtained by the single folding procedure with the complex $G$-matrix interaction and the transition density.
The spin-orbit part of the coupling potential is also obtained in the present single folding procedure.
We apply the MPa $G$-matrix interaction~\cite{YAM13R,YAM14} derived from the realistic nucleon-nucleon interaction, ESC08~\cite{ESC08-1} to the MCC calculation.
The MPa interaction takes into account the three-body force effects with the multi-pomeron exchange potential~\cite{ESC08-1}.
The three-body force induces additional repulsion to the potential and is known to have an important role to improve the analyzing power at forward angles at incident energy from 100 to 200 MeV~\cite{CEG07}.
The MPa interaction has been widely applied not only in the scattering systems~\cite{WWQ15,FUR16,WWQ17,FUR18,FUR19}, but also hypernucler systems~\cite{ISA16,ISA17} and neutron matter~\cite{YAM13R,YAM14,YAM16}.

In this paper, the localized diagonal and coupling potentials are constructed by folding procedure based on Refs.~\cite{KHO02,CEG07}.
The central direct $U^{\rm (CE)}_{D}$ and exchange $U^{\rm (CE)}_{\rm EX}$ potentials are simply described as
\begin{eqnarray}
U^{\rm (CE)}_{D}(R; E_p) &=& \int{ \rho_{tr}(r) v^{\rm (CE)}_{D}(\bm{s}, k_{F}; E_p) d\bm{r} }, \\
U^{\rm (CE)}_{\rm EX}(R; E_p) &=& \int{ \rho_{tr}(x) \frac{3}{k^{\rm (eff)}_F s} j_1(k^{\rm (eff)}_F s) v^{\rm (CE)}_{\rm EX}(\bm{s}, k_{F}; E_p) j_0 (ks) d\bm{r} }, \nonumber \\
\end{eqnarray}
where $R$ is radial distance between incident proton and target nucleus.
$E_p$ is the incident energy.
$\rho_{tr}$ is transition density.
$s$ is radial distance between an incident proton and a nucleon in the target nucleus.
$\bm{s} = \bm{r} - \bm{R}$.
$\bm{x} = \frac{1}{2}(\bm{r} + \bm{R})$.
$k_F$ is the Fermi momentum derived from the densities of the initial and final states.
$j_0$ and $j_1$ are the spherical Bessel function of rank 0 and 1, respectively.
$k_F^{\rm (eff)}$ is a effective Fermi momentum defined in Ref.~\cite{CAM78}.
$v^{\rm (CE)}_D$ and $v^{\rm (CE)}_{\rm EX}$ are the complex $G$-matrix interaction for the central direct and exchange terms, respectively.
The Coulomb potential is also obtained by the folding prescription with the nucleon-nucleon Coulomb interaction and proton density.

The localized diagonal and coupling potentials for the spin-orbit part are also obtained in the same manner as described in Ref.~\cite{CEG07}, while this is the first time to apply to the construction of the coupling potential, as
\begin{eqnarray}
U^{\rm (LS)}_D(R; E_p)&=&\frac{1}{4R^2} \int{\bm{R}\cdot (\bm{R}-\bm{r}) \rho_{tr}(\bm{r}) v^{\rm (LS)}_D(\bm{s}, k_F; E_p) d\bm{r}}, \\
U^{\rm (LS)}_{\rm EX}(R; E_p)&=& \pi \int{ds s^3\left[ \frac{2j_0(ks)}{R}\rho_1 (R, s) + \frac{j_1(ks)}{2k}\delta_0 (R, s) \right] }, \nonumber \\
\end{eqnarray}
where,
%%%%%%%%%%%%%%%%%%%%%%%%%%%%%%%%%%%%%%%%%%%%%%%%%%%%%%%%%%%%%
%\begin{widetext}
\begin{eqnarray}
\delta_0(R, s)&=&\frac{1}{2}\int^{+1}_{-1} {dq \frac{v^{\rm (LS)}_{\rm EX}(s, k_{F}; E_p)}{X} \bigg{[} \frac{3}{k^{\rm{eff}}_{F} s} j_1(k^{\rm{eff}}_{F} s) \frac{d}{dx} \rho_{tr}(x)\bigg|_{x=X}} \nonumber \\
&&+ s\rho_{tr} (X) \frac{d}{dx}k^{\rm{eff}}_{F}(x)\bigg|_{x=X} \frac{d}{dy}\left( \frac{3}{y}j_1(y)\right) \bigg|_{y=k^{\rm{eff}}_{F} s} \bigg] , \\
\rho_1(R, s) &=&\frac{1}{2}\int^{+1}_{-1}dq q v^{\rm (LS)}_{\rm EX}(s, k_{F}; E_p) \frac{3}{k^{\rm{eff}}_{F} s} j_1(k^{\rm{eff}}_{F} s)\rho_{tr} (X),
\end{eqnarray}
%\end{widetext}
%%%%%%%%%%%%%%%%%%%%%%%%%%%%%%%%%%%%%%%%%%%%%%%%%%%%%%%%%%%%%
where $X=\sqrt{R^2+s^2/4+Rsq}$.
$v^{\rm (LS)}_D$ and $v^{\rm (LS)}_{\rm EX}$ are also the complex $G$-matrix interaction for the direct and exchange terms for the spin-orbit interaction, respectively.

In the present calculation, the scattering matrix dependent on the total angular moment is obtained from the folded potentials by solving the coupled-channel equation based on the Stormer method.
The cross section and the analyzing power are calculated with the scattering amplitude derived from the scattering matrix as shown in Refs.~\cite{DNR, TAN78}.

%\vspace{2mm}
%%%%%%%%%%%%%%%%%%%%%%%%%%%%%%%%%%%%%%%%%%%%%%%%%%%%%%%%%%%%%%%%%%%
% results
%%%%%%%%%%%%%%%%%%%%%%%%%%%%%%%%%%%%%%%%%%%%%%%%%%%%%%%%%%%%%%%%%%%
\section{Results}
\label{sec:results}
We here apply the present MCC model to the $p + ^{12}$C elastic and inelastic scatterings.
The relativistic kinematics is used to compute the cross sections and analyzing powers.
We now have the central and spin-orbit potentials as
\begin{eqnarray}
U&=&  U^{\rm (CE)}_D + U^{\rm (CE)}_{\rm EX}
   + (U^{\rm (LS)}_D + U^{\rm (LS)}_{\rm EX}) \bm{\ell} \cdot \bm{\sigma}. \label{eq:d+ex}
\end{eqnarray}
These potentials are complex because the complex $G$-matrix interaction is applied to the folding model calculation.
Therefore, we can rewrite Eq.~(\ref{eq:d+ex}) to
\begin{eqnarray}
U&=&V_{\rm CE} +i W_{\rm CE} + (V_{\rm LS} + i W_{\rm LS}) \bm{\ell} \cdot \bm{\sigma},
\end{eqnarray}
where $V_{\rm CE}$, $W_{\rm CE}$, $V_{\rm LS}$, and $W_{\rm LS}$ are the central real, central imaginary, spin-orbit real, and spin-orbit imaginary potentials, respectively.
Since the complex $G$-matrix is constructed in the infinite nuclear matter, the strength of the imaginary part is often adjusted in the use for the finite nucleus because these level densities are quite different.
Thus, we take the incident-energy-dependent renormalization factor, $N_W = 0.5 + E_p/1000$~\cite{FUR19}, for the imaginary part of the folding model potential as
\begin{eqnarray}
U&=&V_{\rm CE} +i N_W W_{\rm CE} + (V_{\rm LS} + i N_W W_{\rm LS}) \bm{\ell} \cdot \bm{\sigma}.
\end{eqnarray}
Namely, we have no additional parameter to calculate the proton scatterings in this paper.

We take the transition density of the $^{12}$C nucleus from the Algebraic Cluster Model (ACM) calculations and modified ones~\cite{FUN03,FUN06,TAK06}.
We will explain about the modifications of the wave function and transition density just before showing the results.

%\vspace{1mm}
\subsection{Validity of present MCC calculation and role of potentials}

First, we show the validity of the present MCC calculation in the comparison with the experimental data.
Because this is the first time that the spin-orbit part of the coupling potential obtained by the present folding procedure is applied to the proton inelastic scattering.
In addition, it has been confirmed in past phenomenological potentials, but we check the role of the central and spin-orbit parts of the localized diagonal and coupling potentials.

%%%%%%%%%%%%%%%%%%%%%%%%%%%%%%%%%%%%%
% figure
%%%%%%%%%%%%%%%%%%%%%%%%%%%%%%%%%%%%%
\begin{figure}[h]
\centering
\includegraphics[width=10cm]{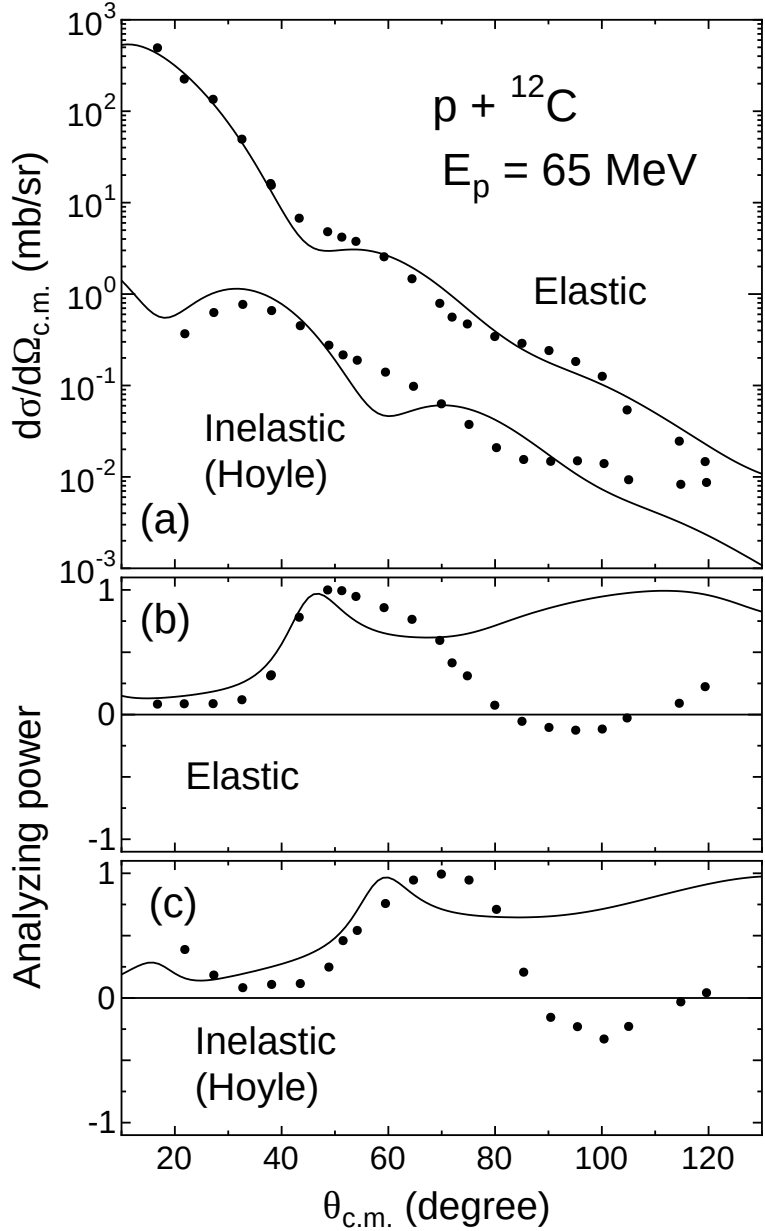}
\caption{\label{fig:65mev}
(a) Elastic and inelastic cross sections, (b) elastic analyzing powers, and (c) inelastic (0$_2^+$) analyzing powers of $p+{}^{12}$C system at 65 MeV.
The solid curves are the results by the present MCC calculation.
The experimental data are taken from \cite{EXFOR,KAT85}.}
\end{figure}
\begin{figure}[h]
\centering
\includegraphics[width=10cm]{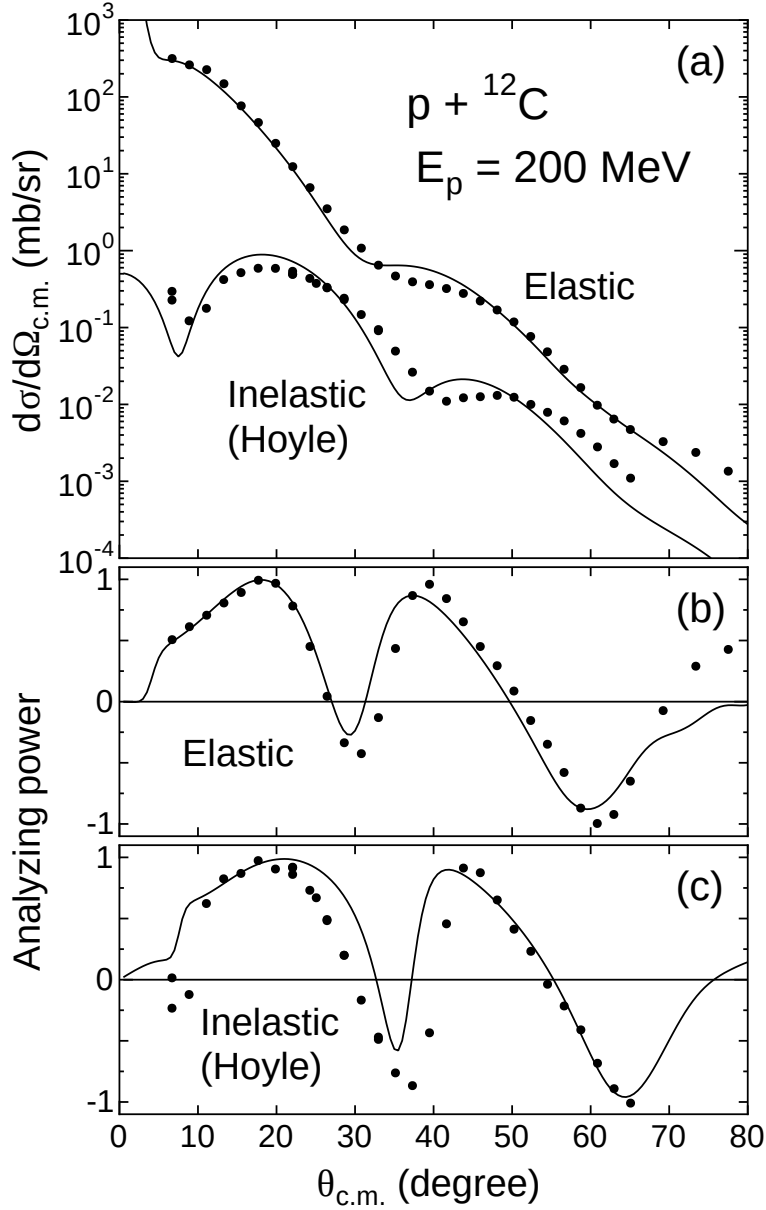}
\caption{\label{fig:200mev}
Same as Fig.~\ref{fig:65mev} but at 200 MeV.
The experimental data are taken from Refs.~\cite{EXFOR,MEY81,COM82}.}
\end{figure}
%%%%%%%%%%%%%%%%%%%%%%%%%%%%%%%%%%%%%

Figures~\ref{fig:65mev} and \ref{fig:200mev} show the elastic and inelastic cross sections and analyzing powers of $p + ^{12}$C system at 65 and 200 MeV.
The calculated results well reproduce the experimental data for the elastic and inelastic cross sections in spite of no adjustable parameter.
We can see the discrepancy between the calculated results and the experimental data only for backward angles of the analyzing powers at 65 MeV. 
Many folding analyses based on the localized and non-localized approaches fail to reproduce the analyzing power of the $p + ^{12}$C system for backward angles~\cite{DOR98,MEL00,CEG07,TOY13,FUR19}.
Therefore, we avoid to discuss for backward angles of the analyzing power at 65 MeV.
We here mention that the channel coupling effect from other excited states (0$^+$, 2$^+$, and 3$^-$) on the elastic and inelastic (0$_2^+$) cross sections is minor, while the test was performed in an approximation with another transition density~\cite{KAM81}.
In this paper, we perform the MCC calculation only with the ground and Hoyle states simply to understand the relation between the transition density and the proton inelastic scattering.

%%%%%%%%%%%%%%%%%%%%%%%%%%%%%%%%%%%%%
% figure
%%%%%%%%%%%%%%%%%%%%%%%%%%%%%%%%%%%%%
\begin{figure}[h]
\centering
\includegraphics[width=10cm]{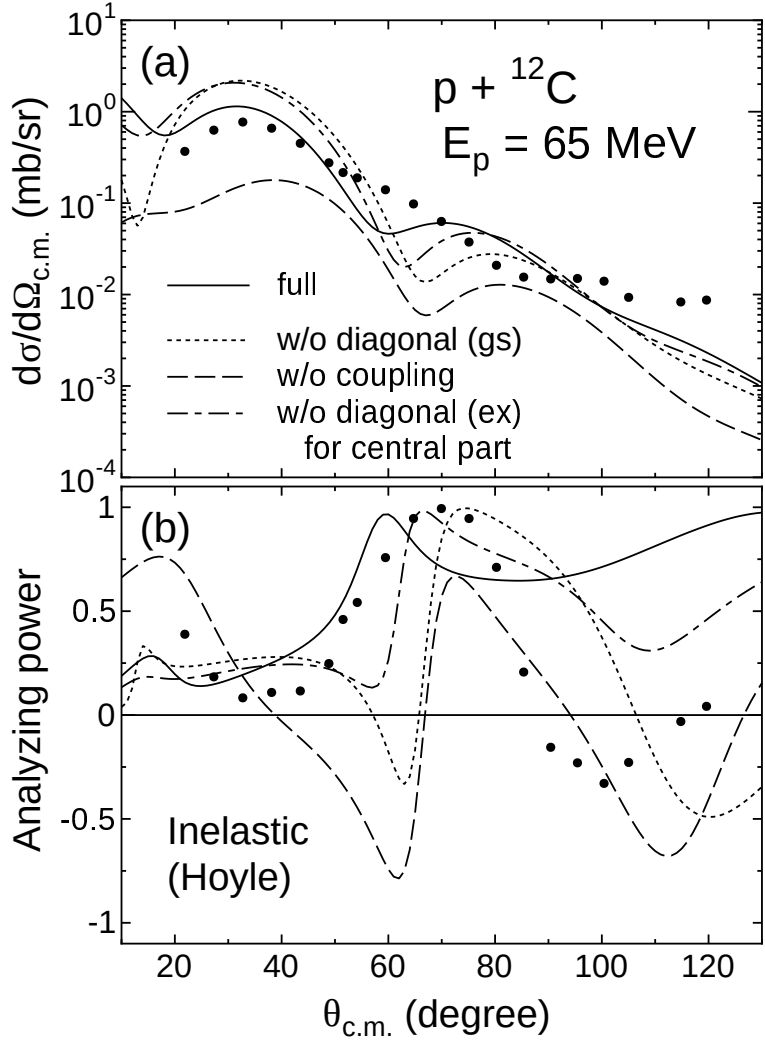}
\caption{\label{fig:65mev-potce}
(a) Inelastic cross sections and (b) inelastic analyzing powers of $p+{}^{12}$C system at 65 MeV.
The solid curves are same results shown in Figs.~\ref{fig:65mev} and \ref{fig:200mev}.
The dotted curves show the results without the diagonal potential of the elastic channel (gs) for the central part.
The dashed curves are obtained without the coupling potential between ground state and excited state (0$^+_2$) for the central part.
The dot-dashed curves show the results without the diagonal potential of the excited channel (0$^+_2$) (ex) for the central part.
}
\end{figure}
\begin{figure}[h]
\centering
\includegraphics[width=10cm]{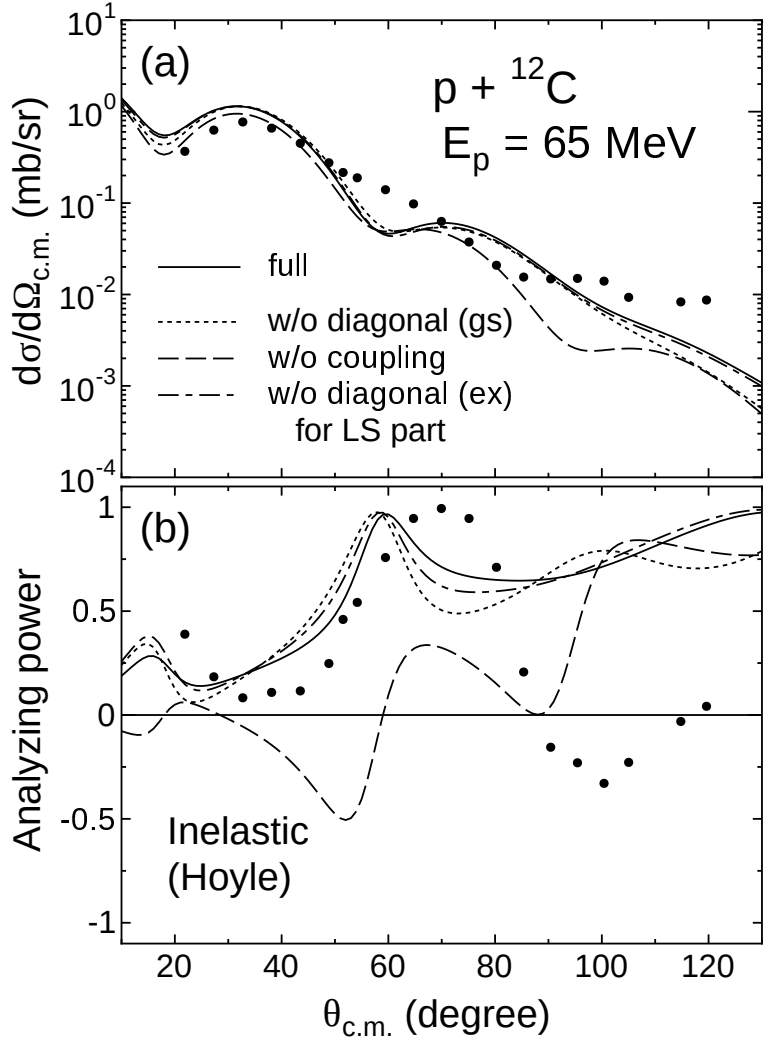}
\caption{\label{fig:65mev-potls}
Same as Fig.~\ref{fig:65mev-potce} but for the spin-orbit (LS) part.}
\end{figure}
\begin{figure}[h]
\centering
\includegraphics[width=10cm]{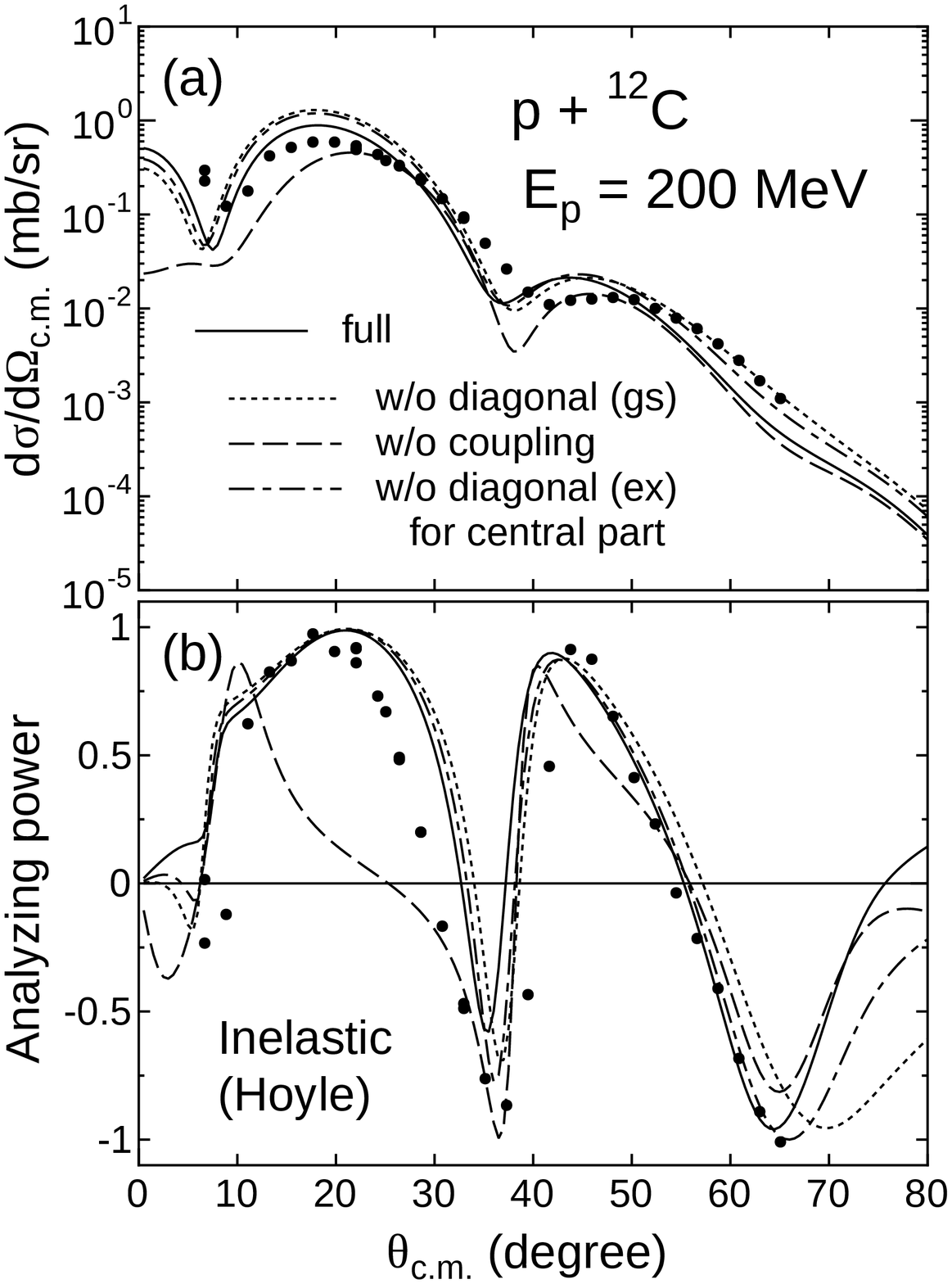}
\caption{\label{fig:200mev-potce}
Same as Fig.~\ref{fig:65mev-potce} but at 200 MeV.}
\end{figure}
\begin{figure}[h]
\centering
\includegraphics[width=10cm]{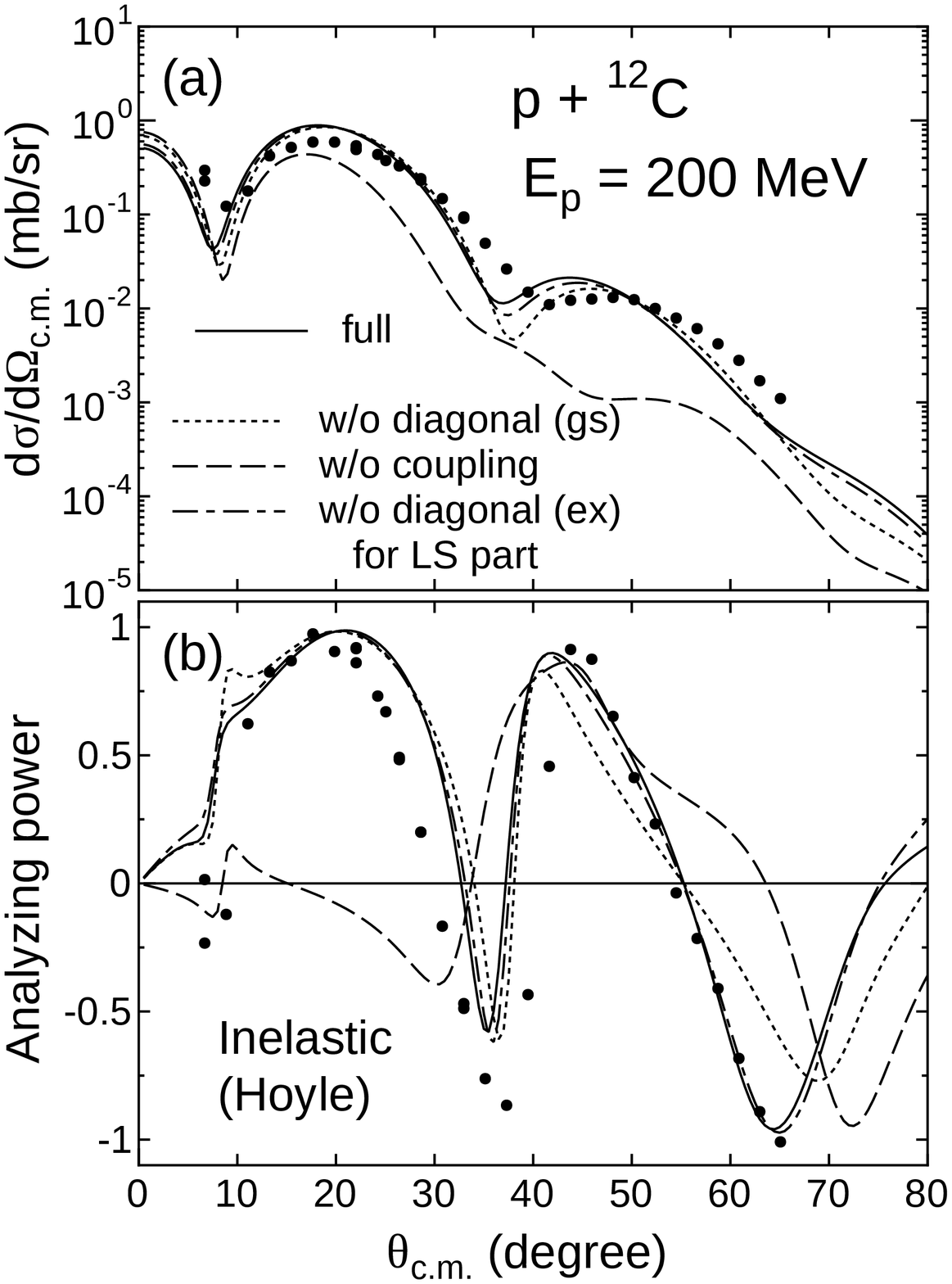}
\caption{\label{fig:200mev-potls}
Same as Fig.~\ref{fig:65mev-potls} but at 200 MeV.}
\end{figure}
%%%%%%%%%%%%%%%%%%%%%%%%%%%%%%%%%%%%%

Next, we check the role of the central and spin-orbit parts of the localized diagonal and coupling potentials obtained by the present folding procedure.
Figures~\ref{fig:65mev-potce}, \ref{fig:65mev-potls}, \ref{fig:200mev-potce}, and \ref{fig:200mev-potls} show the calculated inelastic cross sections and analyzing powers with and without the central and spin-orbit parts of the diagonal and coupling potentials at 65 and 200 MeV, respectively.
The solid curves are same results shown in Figs.~\ref{fig:65mev} or \ref{fig:200mev}.
The dotted curves show the results without the diagonal potential of the elastic channel (gs) for the central or spin-orbit (LS) parts.
The dashed curves are obtained without the coupling potential between ground state and excited state (0$^+_2$) for the central or LS parts.
The dot-dashed curves show the results without the diagonal potential of the excited channel (0$^+_2$) (ex) for the central or LS parts.

In Fig.~\ref{fig:65mev-potce}, the role of central parts is clearly seen in both the inelastic cross section and analyzing power by switching on/off the potentials.
The central part of all potentials is essential to fix the inelastic cross section and analyzing power at 65 MeV.
On the other hand, the role of the spin-orbit part of the coupling potential is clearly seen in Fig.~\ref{fig:65mev-potls}, especially for the inelastic analyzing power.
The spin-orbit part of both the diagonal potentials has a minor role to fix the inelastic cross section and analyzing power at 65 MeV.
The role of the central and spin-orbit parts of the diagonal and coupling potentials obtained by the present folding procedure is consistent with past works~\cite{DNR,SWI72,LIL79,GUS82}, while the target nucleus and the incident energy are different.
At 200 MeV, the central and spin-orbit parts of the coupling potential give a drastic change to the inelastic cross section and the analyzing power as shown in Figs.~\ref{fig:200mev-potce} and \ref{fig:200mev-potls}.
The effect of the diagonal potentials is not so large.
We see the important role of both the central and spin-orbit parts of the coupling potential.
It implies that the inelastic scattering and analyzing power give a property of the transition density because the coupling potentials are derived from the transition density.
In addition, it is simply understood to investigate the contribution of the transition density for proton inelastic scatterings at 200 MeV rather than 65 MeV.

We here note that the same investigation is also performed for the elastic cross section and analyzing power.
However, the diagonal potentials for the elastic channel is just contributed to the elastic cross section and analyzing power.

\subsection{Shape of the transition densities from the proton inelastic scattering}

In this section, we investigate the sensitivity of the transition density to the inelastic cross section and analyzing power.
In the same manner of Ref.~\cite{TAK06}, we apply the modified wave function and the modified transition density to the present MCC calculation, respectively.
Namely, we perform the investigation with the artificial drastic change rather than fine structural change.

\subsubsection{Sensitivity for the size of the excited state and the strength of the transition density}

%%%%%%%%%%%%%%%%%%%%%%%%%%%%%%%%%%%%%
% figure
%%%%%%%%%%%%%%%%%%%%%%%%%%%%%%%%%%%%%
\begin{figure}[h]
\centering
\includegraphics[width=10cm]{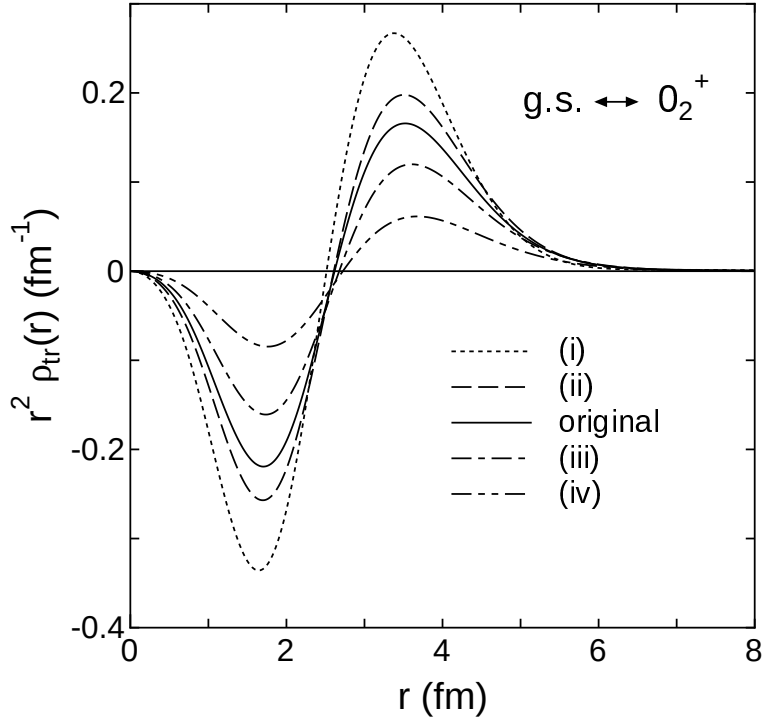}
\caption{\label{fig:trdens-r}
Transition densities obtained by modified ACM wave functions.
The meaning of the curves is introduced in the text.}
\end{figure}
%%%%%%%%%%%%%%%%%%%%%%%%%%%%%%%%%%%%%

First, we introduce the transition density based on the modified wave function presented in Ref.~\cite{TAK06}.
There are four types of modified ACM wave functions (i)--(iv), the root-mean-square (rms) radii $<r^2>^{1/2}$ of which are (i) 2.97 fm, (ii) 3.55 fm, (iii) 4.38 fm, and (iv) 5.65 fm, while that of the original ACM wave function is 3.81 fm~\cite{FUN03}.
Again, it should be noted that the orthogonality of the 0$^+_2$ state and the 0$^+_1$ ground state wave functions is satisfied also in the modified ACM calculation.
In the present folding model, just the transition density is needed.
However, the transition density is constructed with the modified wave function.
Therefore, we use the expression as the modified ACM wave functions in this paper.
The transition density between the ground and 0$^+_2$ states is calculated from the ground state wave function and each of these wave functions for the 0$^+_2$ state.
The ground state wave function is not modified.
Figure~\ref{fig:trdens-r} shows the transition densities obtained by the modified ACM wave function which is also shown in Ref.~\cite{TAK06}.
Because the coupling potentials have an important role to fix the inelastic (0$^+_2$) scatterings as shown in Figs~\ref{fig:65mev-potce}, \ref{fig:65mev-potls}, \ref{fig:200mev-potce}, and \ref{fig:200mev-potls}, we here show the transition density.

%%%%%%%%%%%%%%%%%%%%%%%%%%%%%%%%%%%%%
% figure
%%%%%%%%%%%%%%%%%%%%%%%%%%%%%%%%%%%%%
\begin{figure}[h]
\centering
\includegraphics[width=10cm]{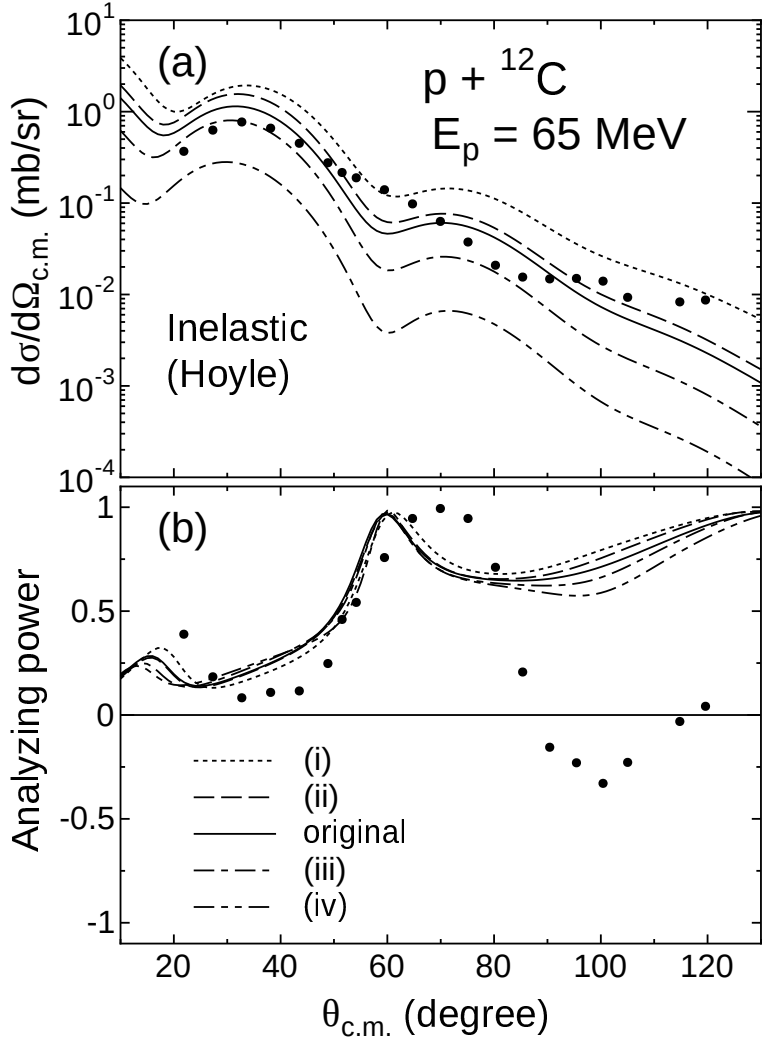}
\caption{\label{fig:65mev-r}
(a) Inelastic cross sections and (b) inelastic analyzing powers of $p+{}^{12}$C system at 65 MeV with the transition densities obtained by the modified ACM wave function.
}
\end{figure}
\begin{figure}[h]
\centering
\includegraphics[width=10cm]{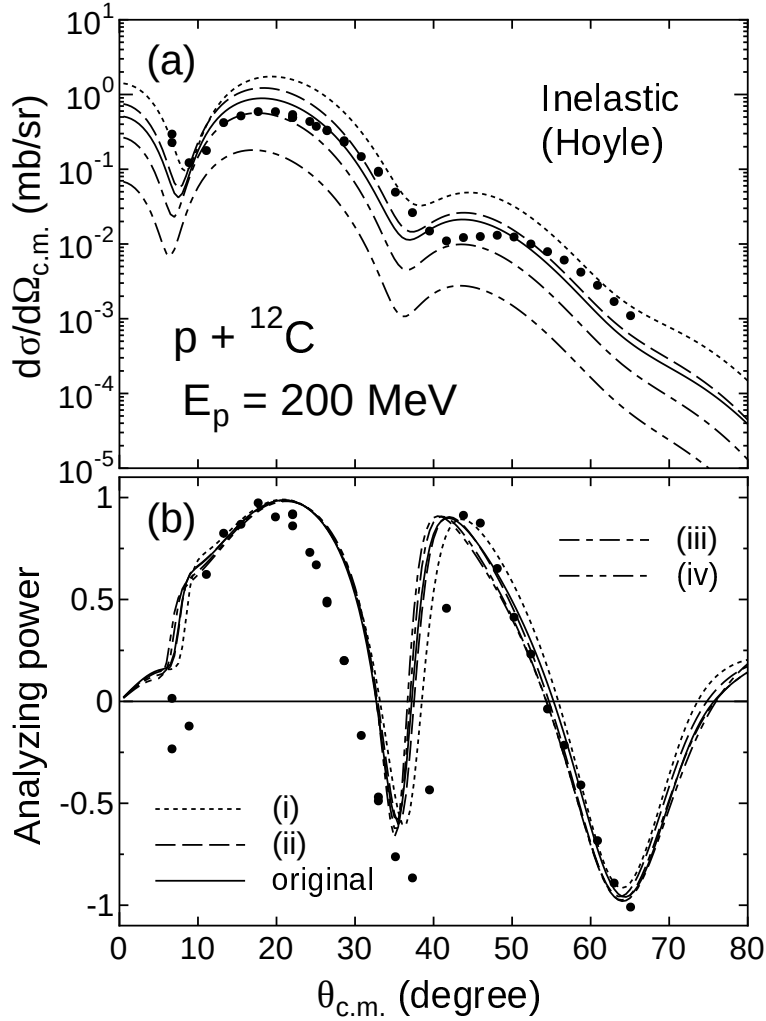}
\caption{\label{fig:200mev-r}
Same as Fig.~\ref{fig:65mev-r} but at 200 MeV.}
\end{figure}
%%%%%%%%%%%%%%%%%%%%%%%%%%%%%%%%%%%%%

Figures~\ref{fig:65mev-r} and \ref{fig:200mev-r} show the calculated inelastic cross section and analyzing power with the transition densities by the modified ACM wave function.
For the inelastic cross section, the strength of the cross section is drastically changed at 65 and 200 MeV.
However, the phase of the diffraction pattern for the angle direction is not changed.
This result is completely comparable to the $\alpha$ inelastic scattering obtained by one of authors and his collaborator~\cite{TAK06}.
In addition, this result is simply understood by the relation between the inelastic form factor and the transition density through Fourier transform.
We confirmed that the size of the excited state has no effect on the phase of the diffraction pattern for the angle direction.
The size of the excited state changes the strength of the transition density and the absolute value of the inelastic cross section.
Next, we focus on the analyzing power.
Not only the size of the excited state but also the strength of the transition density has no effect on the inelastic analyzing power.
This result implies that the inelastic analyzing power tell us the shape of the transition density independent of the transition strength.
We will discuss such situation in the next section.

\subsubsection{Sensitivity for the shape of the transition density}

%%%%%%%%%%%%%%%%%%%%%%%%%%%%%%%%%%%%%
% figure
%%%%%%%%%%%%%%%%%%%%%%%%%%%%%%%%%%%%%
\begin{figure}[h]
\centering
\includegraphics[width=10cm]{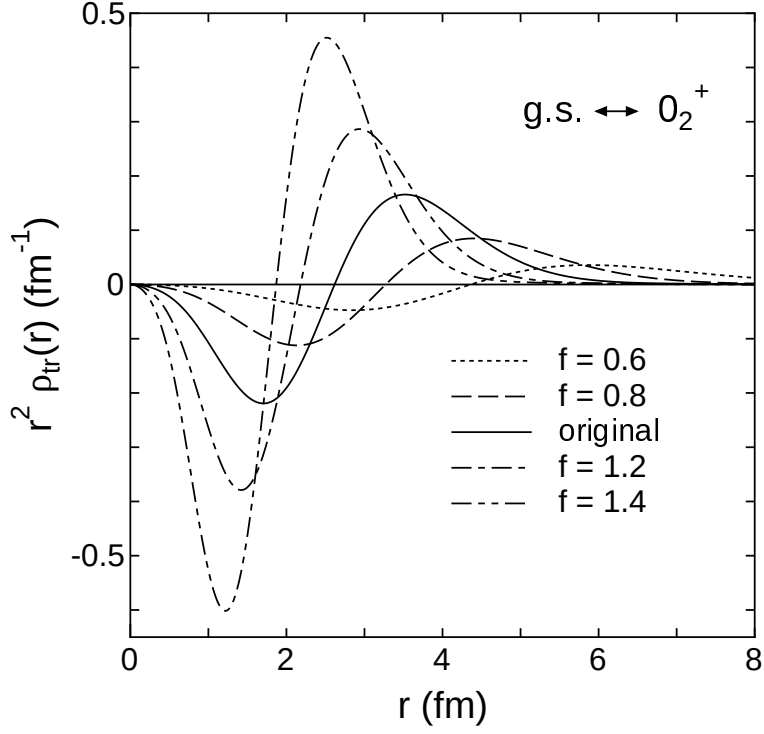}
\caption{\label{fig:trdens-f}
Modified transition densities.
The detail is introduced in the text.}
\end{figure}
%%%%%%%%%%%%%%%%%%%%%%%%%%%%%%%%%%%%%

Next, we investigate the sensitivity for the shape of the transition density by the artificial modification of the transition density.
According to Ref.~\cite{TAK06}, the modified transition density is obtained as
\begin{equation}
\rho'_{tr}(r) = N \rho_{tr}(fr).
\end{equation}
Here, $\rho_{tr}$ is the original ACM transition density.
$f$ and $N$ are scaling and normalization factors, respectively.
$N$ is fixed to keep the $r^2$ moment of the transition density to the value evaluated from the original transition density.
Figure~\ref{fig:trdens-f} shows the modified transition densities which is also shown in Ref.~\cite{TAK06}.
By the modification, the shape of the transition density is drastically changed.
Again, we mention that the wave function of the excited (0$_2^+$) state is fixed to be original ACM wave function.

%%%%%%%%%%%%%%%%%%%%%%%%%%%%%%%%%%%%%
% figure
%%%%%%%%%%%%%%%%%%%%%%%%%%%%%%%%%%%%%
\begin{figure}[h]
\centering
\includegraphics[width=10cm]{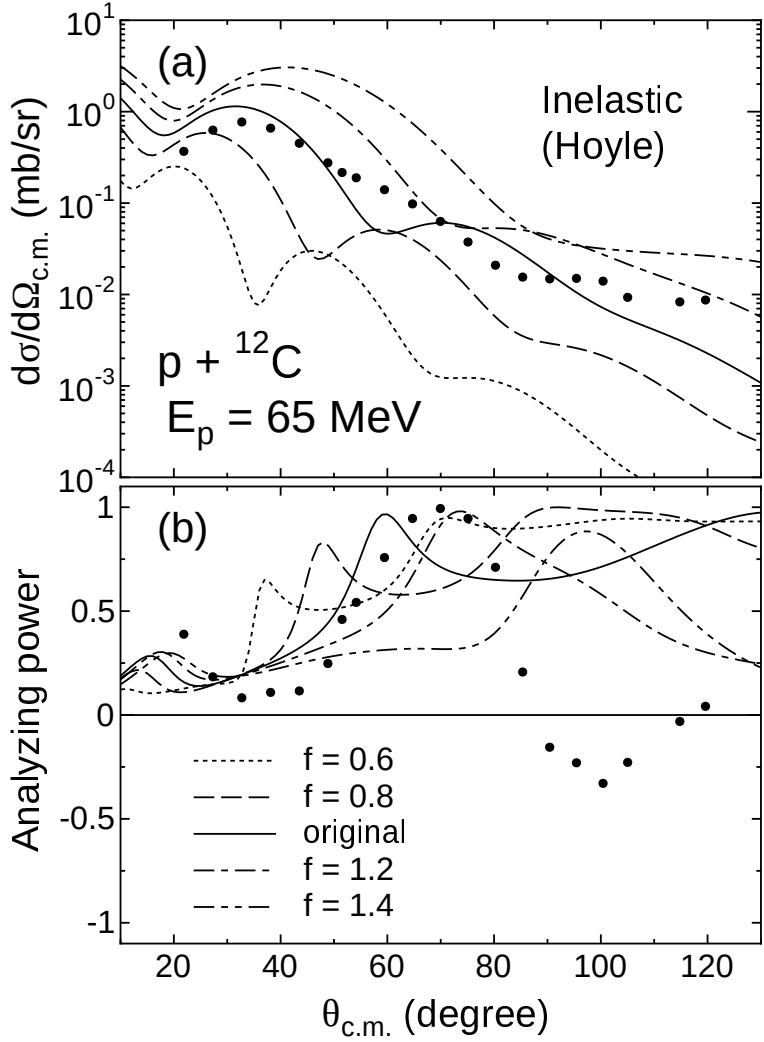}
\caption{\label{fig:65mev-f}
Same as Fig.~\ref{fig:65mev-r} but with the transition densities as shown in Fig.~\ref{fig:trdens-f}.}
\end{figure}
\begin{figure}[h]
\centering
\includegraphics[width=10cm]{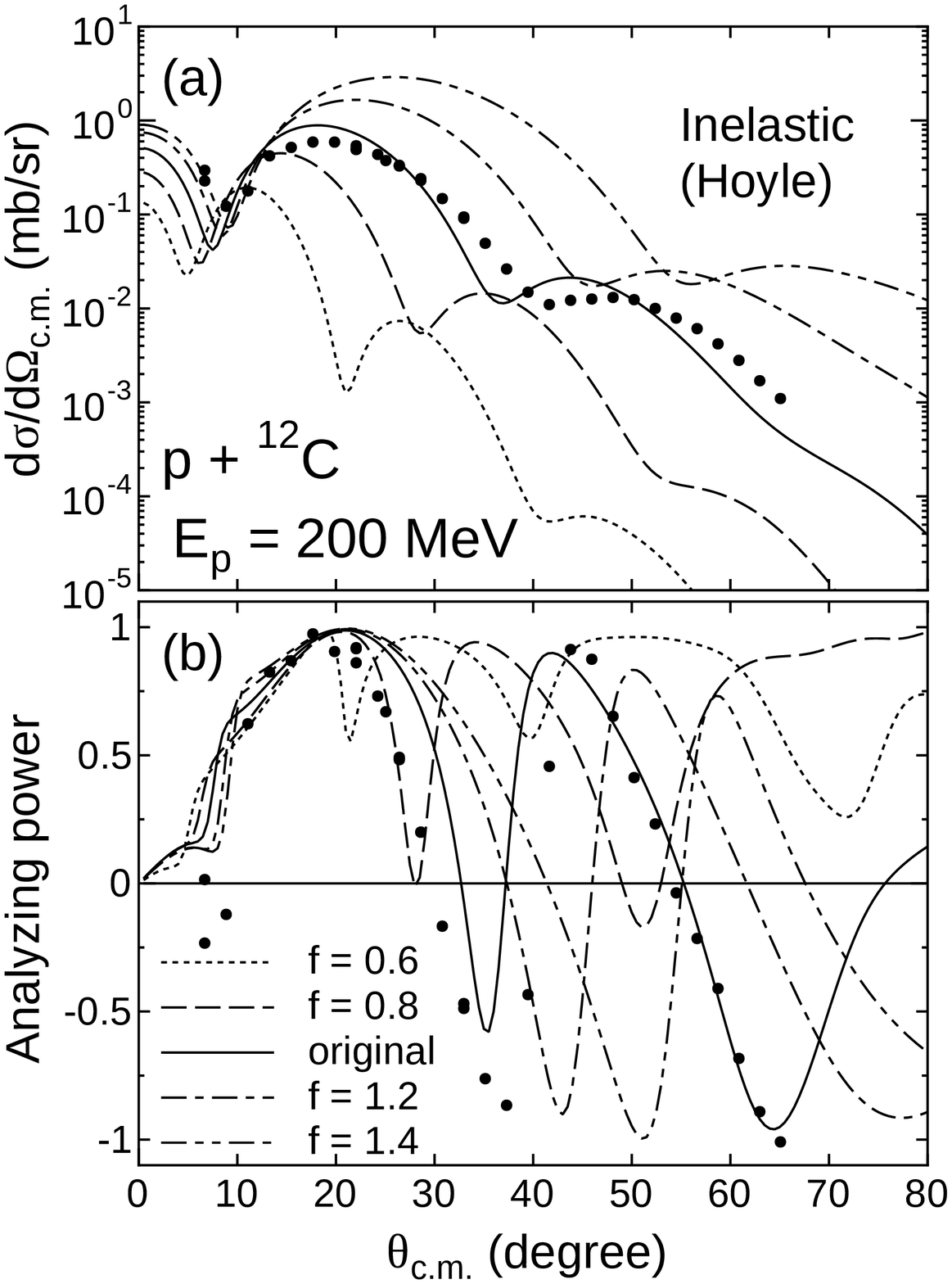}
\caption{\label{fig:200mev-f}
Same as Fig.~\ref{fig:65mev-f} but at 200 MeV.}
\end{figure}
%%%%%%%%%%%%%%%%%%%%%%%%%%%%%%%%%%%%%

Figures~\ref{fig:65mev-f} and \ref{fig:200mev-f} show the calculated inelastic cross section and analyzing power with the transition densities as shown in Fig.~\ref{fig:trdens-f}.
The calculated inelastic cross sections are drastically changed.
Especially, the phase of the diffraction pattern for the angle direction is shifted.
This result is also comparable to the $\alpha$ inelastic scattering obtained by one of authors and his collaborator~\cite{TAK06}.
Again, it is simply understood by the relation between the inelastic form factor and the transition density.
Contrary to the results as shown in Figs.~\ref{fig:65mev-r} and \ref{fig:200mev-r}, the calculated analyzing powers show the drastic change at both incident energies.
The shift of angular direction for the analyzing power is comparable with that for the cross section.
Both the inelastic cross section and analyzing power tell us the properties (the strength and shape) of transition density.

%\vspace{2mm}
%%%%%%%%%%%%%%%%%%%%%%%%%%%%%%%%%%%%%%%%%%%%%%%%%%%%%%%%%%%%%%%%%%%
% summary
%%%%%%%%%%%%%%%%%%%%%%%%%%%%%%%%%%%%%%%%%%%%%%%%%%%%%%%%%%%%%%%%%%%
\section{Summary}
\label{sec:summary}

We have constructed a microscopic coupled channel (MCC) calculation for proton elastic and inelastic scatterings.
The localized diagonal and coupling potentials including the spin-orbit part is obtained by folding the complex G-matrix interaction with the transition density.
This is the first time that the present folding prescription for the spin-orbit part has been applied to the proton inelastic scattering, while for the monopole transition only.
The proton elastic and inelastic (0$^+_2$) cross sections and analyzing powers are calculated by the $^{12}$C target at 65 and 200 MeV.
The calculated cross section and analyzing power well reproduce the experimental data.
Namely, the present folding prescription gives the suitable central and spin-orbit parts of the localized diagonal and coupling potentials without an ambiguity.
The role of the diagonal and coupling potentials for the central and spin-orbit parts is checked by switching on/off the potentials.
The central and spin-orbit parts of the coupling potential has an important role to fix the inelastic cross section and analyzing power at 65 and 200 MeV.

We apply the modified wave function and the modified transition density to the MCC calculation to investigate the relation between the transition density and the proton inelastic scattering.
This is the merit of the localized approach because the information as the transition density obtained in the middle of the theoretical calculation can be changed artificially.
The contribution of the strength and shape of the transition density is clearly seen in the inelastic cross section.
However, this result is simply understood by the relation between the inelastic form factor and the transition density through Fourier transform.
On the other hand, the strength of the transition density has no effect on the inelastic analyzing power.
The inelastic analyzing power is sensitive only to the shape of the transition density.
Finally, we make clear the property of the inelastic analyzing power derived from the transition density without an ambiguity.
For the proton scatterings, both the inelastic cross section and analyzing power have important role to investigate the properties of the transition density.
Although only monopole transition was treated in this study, the present framework will be extended to include other transitions in the future.

\vspace{2mm}
%%%%%%%%%%%%%%%%%%%%%%%%%%%%%%%%%%%%%%%%%%%%%%%%%%%%%%%%%%%%%%%%%%%
% acknowledgments
%%%%%%%%%%%%%%%%%%%%%%%%%%%%%%%%%%%%%%%%%%%%%%%%%%%%%%%%%%%%%%%%%%%
%\section*{Acknowledgment}
\acknowledgments
This work was supported by Japan Society for the Promotion of Science (JSPS) KAKENHI Grant Number JP20K03944.
%JP20K03944: kiban C (Furumoto)

%%%%%%%%%%%%%%%%%%%%%%%%%%%%%%%%%%%%%%%%%%%%%%%%%%%%%%%%%%%%%%%%%%%
% references
%%%%%%%%%%%%%%%%%%%%%%%%%%%%%%%%%%%%%%%%%%%%%%%%%%%%%%%%%%%%%%%%%%%
%\newpage %Just because of unusual number of tables stacked at end
%\bibliography{FT-C12+p-PRC}% Produces the bibliography via BibTeX.

\end{document}